\documentclass[aps,showpacs]{revtex4}
\usepackage{psfig}

\def\order#1{{\cal O}\!\left(#1\right)}
\newcommand{\ba}{\begin{eqnarray}}
\newcommand{\ea}{\end{eqnarray}}
\newcommand{\be}{\begin{equation}}
\newcommand{\ee}{\end{equation}}
\newcommand{\mm}{\omega}

\begin{document}

\title{
%
%
\[ \vspace{-2cm} \]
\noindent\hfill\hbox{\rm  } \vskip 1pt \noindent\hfill\hbox{\rm
Alberta Thy 04-04}
\vskip 10pt
Heavy-to-light decays with a two-loop accuracy
}

\author{Ian Blokland, Andrzej Czarnecki, and Maciej \'Slusarczyk}
\affiliation{
Department of Physics, University of Alberta\\
Edmonton, AB\ \  T6G 2J1, Canada}

\author{Fyodor Tkachov}
\affiliation{
Institute for Nuclear Research, Russian Academy of Sciences\\
Moscow, 117312, Russian Federation}

\begin{abstract}
We present a determination of a new class of three-loop Feynman
diagrams describing heavy-to-light transitions.  We apply it to
find the $\order{\alpha_s^2}$ corrections to the top quark decay
$t\to bW$ and to the distribution of lepton invariant mass in the
semileptonic $b$ quark decay $b\to ul\nu$.  We also confirm the
previously determined total rate of that process as well as the
$\order{\alpha^2}$ corrections to the muon lifetime.
\end{abstract}


\maketitle

The determination of higher order corrections in perturbative quantum
field theory is notoriously difficult, and with the general
tendency towards precision measurements in particle physics, each
newly-won class of perturbative integrals expands the possibilities
for phenomenological analyses. For instance, quantum corrections
to decays of neutral particles, such as a virtual photon or a $Z$
boson into hadrons, are known to sixth order in perturbation
theory, $\order{\alpha_s^3}$, and even some $\order{\alpha_s^4}$
effects have been studied.  Those results have been very useful in
determining a variety of Standard Model parameters such as the
$Z$ boson properties, the strong coupling constant, and the
running of the electromagnetic coupling constant
\cite{Steinhauser:2002rq}.

Much less is known about radiative corrections to processes with a
charged particle in the initial state.  Only relatively recently
have first results been obtained in fourth order perturbation
theory, $\order{\alpha_s^2}$ and $\order{\alpha^2}$, primarily for
total decay rates.  The technical challenge in such calculations
is the presence of massive propagators.  For example, consider the
muon decay.  Since the muon is charged, it can emit  photons, and
the resulting amplitudes will involve propagators of a virtual
muon.  Its mass sets the energy scale of the process and cannot be
treated as a small parameter.

\begin{figure}[htb]
\hspace*{0mm}\psfig{figure=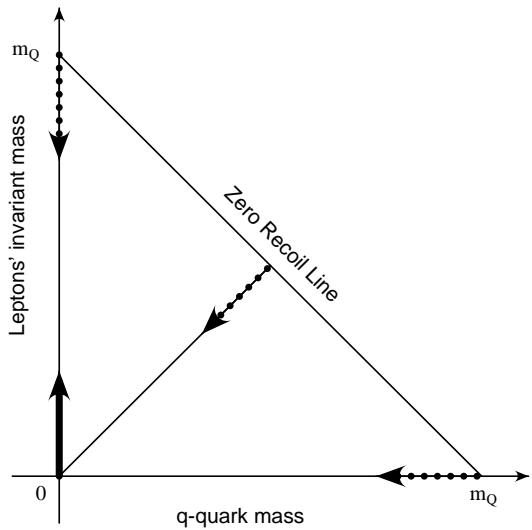,width=70mm}
\caption{Kinematical boundaries of the semileptonic decays $Q\to q+
\mbox{leptons}$. The solid arrow shows the expansion presented in
this paper.  Previously known expansions are indicated with dotted
arrows.} \label{fig:tri}
\end{figure}

The presence of massive propagators is an obstacle in evaluating
the multi-loop diagrams required by precise measurements of heavy
quark and lepton decays. So far, genuine $\order{\alpha_s^2}$
corrections to heavy quark decays are known only for semileptonic
processes, $Q \to q l\nu(gg)$, and only for some kinematic cases.
One approach that has been successful consists in expanding
Feynman diagrams around the zero recoil limit: when the quark $q$
remains at rest with respect to $Q$.  The kinematics of
semileptonic decays can be represented by a triangle, since the
invariant mass of the leptons together with the mass of the final
state quark $q$ cannot exceed the mass of the decaying quark. This
is depicted in Fig.~\ref{fig:tri}.
The diagonal boundary corresponds to the zero recoil limit, in
which the $\order{\alpha_s^2}$ effects are known
\cite{zerorecoil,zerorecoilA,Franzkowski:1997vg}.  Also shown are
the starting points of previously studied expansions.  Those
results have helped improve the knowledge of the $b$ quark
lifetime and the determination of the CKM matrix element $V_{cb}$.

However, the expansion around zero recoil converges slowly near
the origin in Fig.~\ref{fig:tri}, that is, when both the quark $q$ and
the lepton pair are light; in this case computations become
prohibitively expensive. Two other approaches have been used in
such cases. First, for the phenomenologically important decays
$\mu\to e\nu\bar\nu$ and $b\to u l\nu$, the total lifetimes have
been determined in \cite{vanRitbergen:1998yd,vanRitbergen:1999gs}
by analytically calculating imaginary parts of four-loop diagrams.
That heroic effort is difficult to extend, for example, to differential
distributions. The other approach consists in expanding diagrams
in an artificial parameter, for example the ratio of muon masses
outside and inside loops, and using Pad\'e approximants to sum the
expansion for the physical value of this parameter. This approach
was used to check the muon and $b\to u$ results
\cite{Steinhauser:1999bx,Chetyrkin:1999ju}. It yields results
scattered around the exact values which are often sufficient for
applications. The drawback of this method is that it is very
difficult to estimate the errors reliably.

The purpose of this study is to extend the method of expansions
beyond the zero recoil limit. We start directly at the origin of
Fig.~\ref{fig:tri}, which corresponds to the kinematics of a top
quark decay into a massless $b$ quark and a massless $W$ boson. We
then treat the $W$ mass as a small perturbation and compute
several terms of the resulting expansion.  We stop when we can
smoothly match to the previously obtained expansion around the case
of the $W$ boson equally heavy as the decaying quark
\cite{Czarnecki:2001cz}.  The physical top quark decay corresponds
to a specific value of the $W$ mass, but in the more general decay
$Q\to q+ \mbox{leptons}$, the term ``$W$ mass'' refers to the
invariant mass of the lepton pair, and it is in this context that
we can relate $t\rightarrow bW$ to $b\rightarrow ul\nu$ and $\mu
\rightarrow e\nu \overline{\nu}$.

This is the first time that exact results are available in this
limit and we can now address a number of interesting problems.  We
obtain an accurate value of the $\order{\alpha_s^2}$ correction to
the top quark lifetime.  The combination of our results with the
expansion around the heavy $W$ case allows us to give a complete
description of the differential distribution of $b\to ul\nu$ decay
in the invariant mass of leptons, and thus improves the
theoretical description of this decay, important for the
determination of $V_{ub}$.  We also check  the muon and $b\to u$
lifetime corrections with a relative error of about $2\times
10^{-4}$. In the future, the same method can be employed to
improve perturbative corrections to mixing processes such us
$B_d\leftrightarrow \overline B_d$ and $B_s\leftrightarrow
\overline B_s$.

\begin{figure}[htb]
\begin{tabular}{ccc}
\psfig{figure=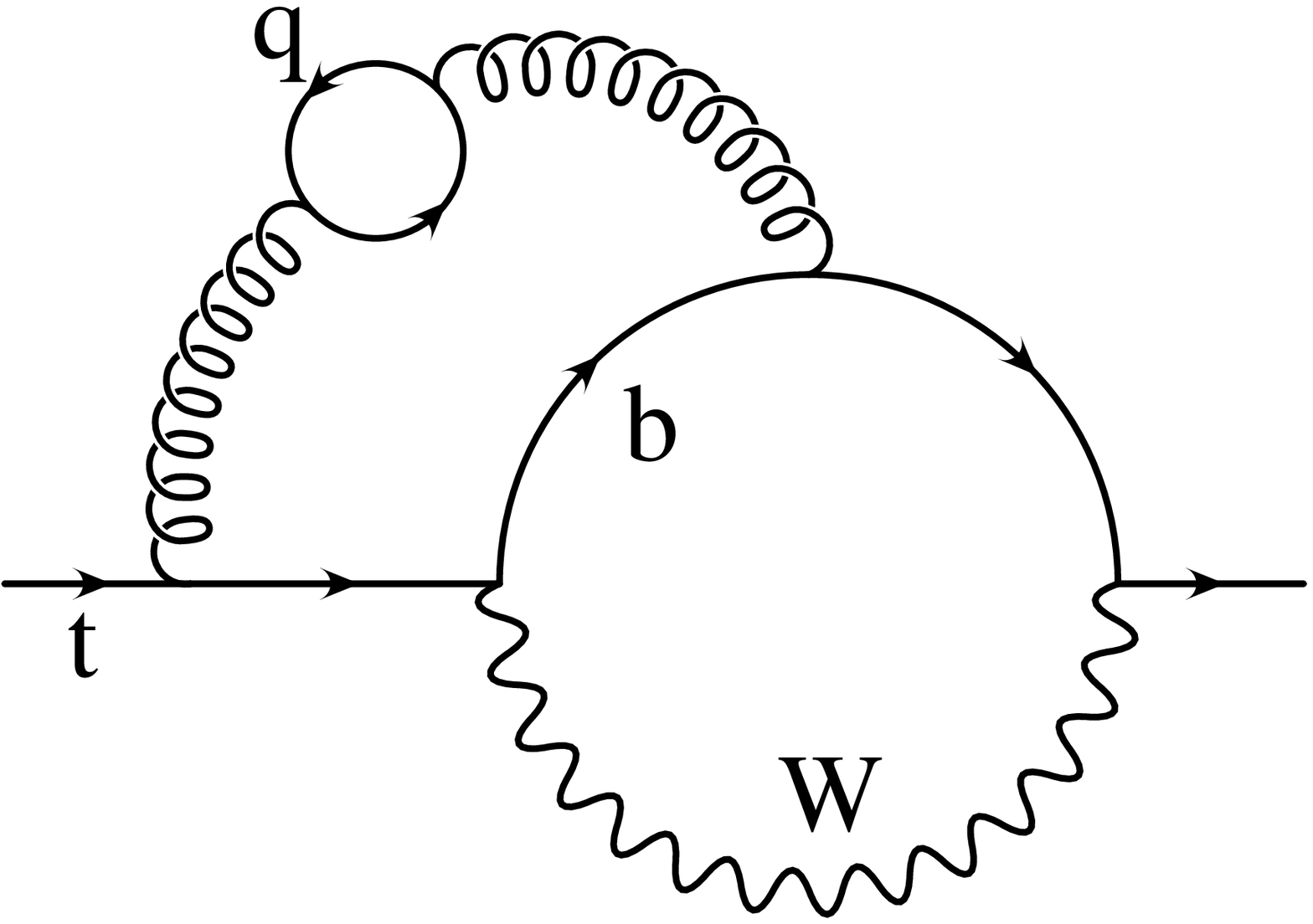,width=42mm}
&
\hspace{5mm}
\psfig{figure=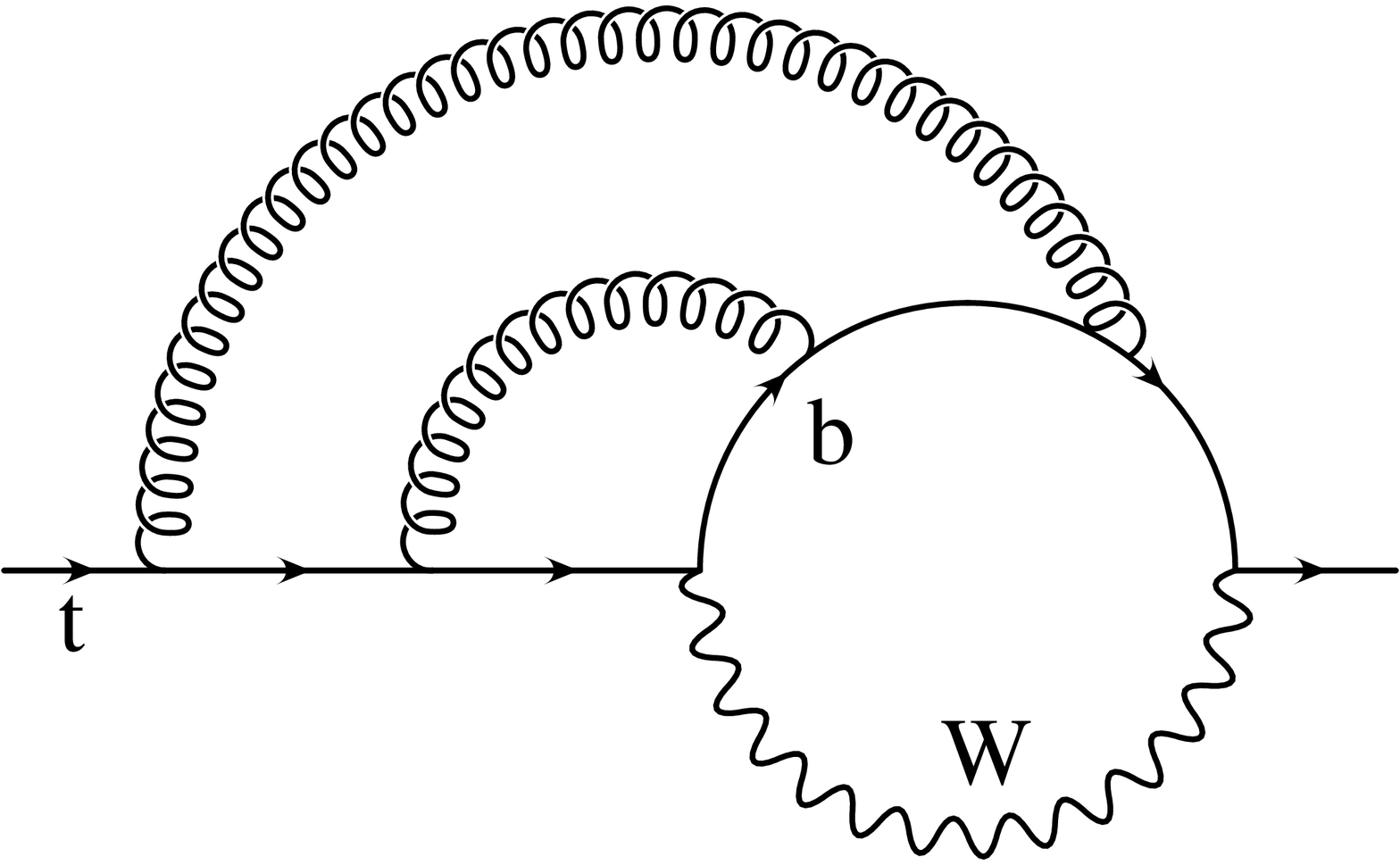,width=52mm}
\hspace{5mm}
&
\psfig{figure=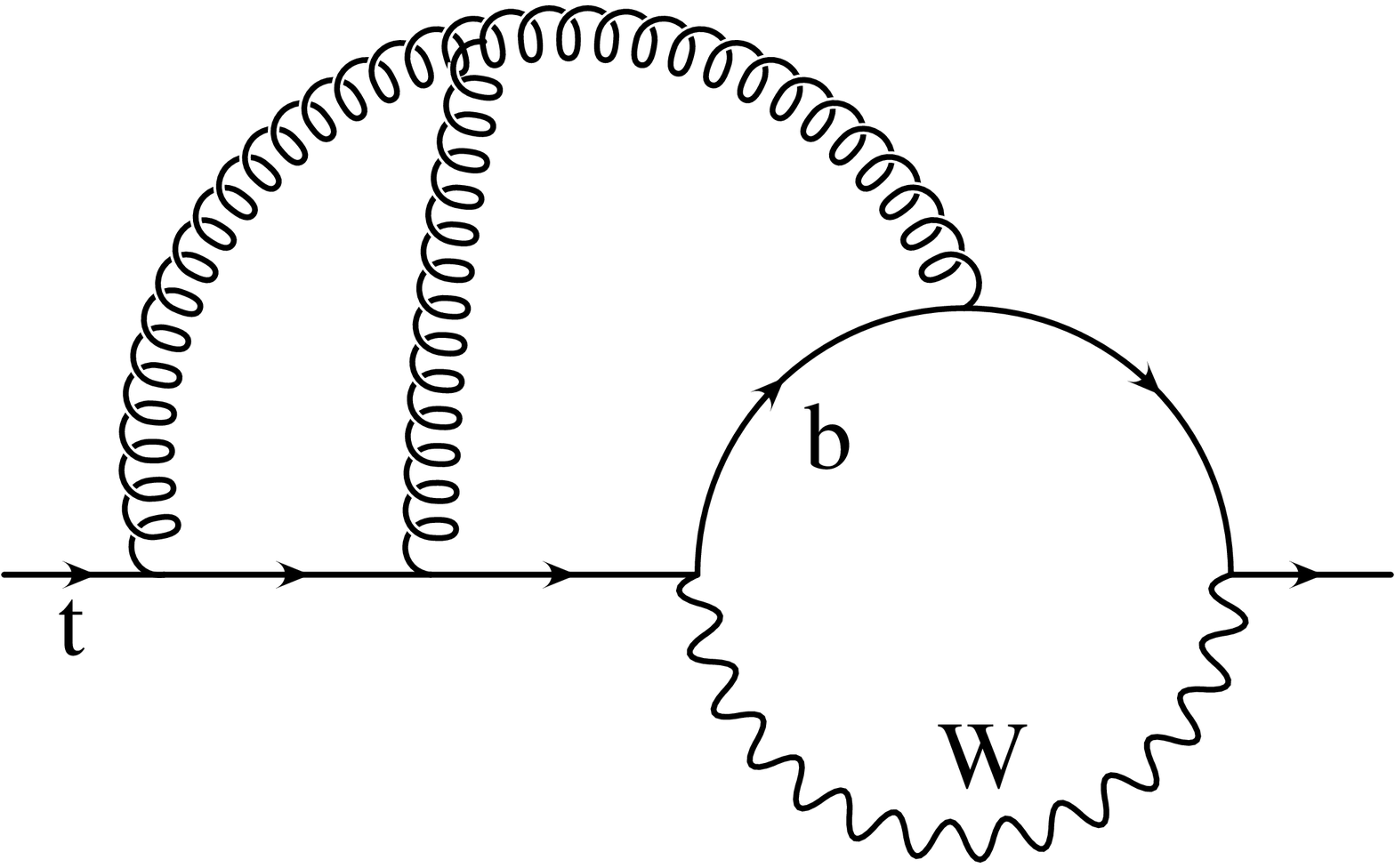,width=52mm}
\\
(a) & (b) & (c)
\end{tabular}
\caption{Examples of diagrams whose cuts contribute to the
$\mathcal{O}(\alpha_s^2)$ decay rate for $t\rightarrow bW$: (a)
light or heavy quarks; (b) abelian; (c) non-abelian.}
\label{fig:diagrams}
\end{figure}

In Fig.~\ref{fig:diagrams} we show three examples of the diagrams
that we have to consider in order to calculate $t\rightarrow bW$
at $\order{\alpha_s^2}$. We use the optical theorem to connect the
imaginary parts of such diagrams with contributions to the decay.
Note that we customarily speak about two-loop corrections when
what we actually need to compute are the imaginary parts of
three-loop diagrams.  The various cuts correspond to two-loop
virtual corrections or emissions of one or two real quanta.

With $m_b=0$, there are two scales in the problem: $m_t$ and
$m_W$.  We define an expansion parameter $\mm=m_W^2/m_t^2$ so that
the two scales can be expressed as hard and soft ($\order{1}$ and
$\order{\sqrt{\mm}}$, using $m_t$ as the unit of energy).
Contributions arising from these two scales are identified using
asymptotic expansions so that we must consider two regions.  In
the first region, all the loop momenta are hard and the $W$
propagator can be expanded as a series, in powers of $\mm$, of
massless propagators.  In the second region, the gluon momenta are
hard but the loop momentum flowing through the $W$ is soft. In
this region, the diagrams factor into a product of a two-loop
self-energy type integral and a one-loop vacuum bubble integral
with a scale of $m_W$. The leading contribution from this second
region is $\order{\mm^2}$, and the interplay between the two
regions gives rise to terms with a large logarithm $\ln \mm$.

All scalar integrals arising in the problem can be expressed in
terms of $9$ basic topologies. We use differential-algebraic
identities to reduce all loop integrals in both regions to a
combination of $24$ master integrals.  The resulting large linear
systems can be solved in a few ways. In the traditional method
\cite{Tkachov:1981wb}, one inspects the structure of the identities
and rearranges them manually into the form of recurrence relations
for an efficient iterative solution of the system. This ``by
inspection'' method has proven to be very successful in numerous
applications (e.g., \cite{bro91a, vanRitbergen:1998yd,
vanRitbergen:1999gs, Czarnecki:2001cz}) but it requires much human
work to implement. Conversely, a straightforward
solution of the linear system is much more expensive computationally and
was first achieved only recently
\cite{Laporta:2001dd}. In our calculation \cite{csc} we used the traditional approach
(programmed in FORM \cite{Vermaseren:2000nd}) as well as a
modified version of the new algorithm for which we implemented
a dedicated computer algebra system. In both cases we independently
obtained identical results which serve as a check of correctness
but also enable us to compare these two methods. Details of the
implementation of both methods and the evaluation of master
integrals will be presented in a forthcoming technical paper.

The final result for the top quark decay width can be written as
\be \Gamma(t\rightarrow bW) = \Gamma_0 \left[ X_0 +
\frac{\alpha_s}{\pi} X_1 + \left( \frac{\alpha_s}{\pi} \right)^2
X_2 \right], \qquad \Gamma_0 \equiv \frac{G_F m_t^3 \left| V_{tb}
\right|^2}{8\sqrt{2}\pi} \ . \ee Throughout this paper, we use
$\alpha_s \equiv \alpha_s^{\overline{MS}}(\mu)$, where $\mu$ is
the pole mass of the decaying quark.  The tree-level and
$\order{\alpha_s}$ coefficients  are already known analytically
\cite{jk2}, \ba X_0 &=& 1 - 3\mm^2 + 2\mm^3 \ ,
\\
X_1 & = & C_F \left[ \left( \frac{5}{4} - \frac{\pi^2}{3} \right)
+ \frac{3}{2}\mm + \mm^2 \left( \pi^2 - 6 + \frac{3}{2}L \right)
+\order{\mm^3 L}\right], \qquad  L\equiv \ln \mm. \ea The
$\order{\alpha_s^2}$ result can be subdivided into four
gauge-invariant pieces, \be X_2 = C_F \left( T_R N_L X_L + T_R N_H
X_H + C_F X_A + C_A X_{NA} \right) \ , \ee where $C_F=4/3$,
$C_A=3$, and $T_R=1/2$ are the usual SU(3) color factors and $N_L$
and $N_H$ denote the number of light $(m_q=0)$ and heavy
$(m_q=m_t)$ quark species.  For the coefficients $X_L$, $X_H$,
$X_A$, and $X_{NA}$, we have obtained a series to at least
$\mm^5$, of which the leading terms are \ba X_L &\simeq& \left[
-\frac{4}{9} + \frac{23\pi^2}{108} + \zeta_3 \right] + \mm \left[
-\frac{19}{6} + \frac{2\pi^2}{9} \right] + \mm^2 \left[
\frac{745}{72} - \frac{31\pi^2}{36} - 3\zeta_3 - \frac{7}{4}L
\right] \ ,
\nonumber \\
X_H & \simeq & \left[
\frac{12991}{1296} - \frac{53\pi^2}{54} - \frac{1}{3}\zeta_3
\right] + \mm \left[ -\frac{35}{108} - \frac{4\pi^2}{9} + 4\zeta_3
\right] + \mm^2 \left[ -\frac{6377}{432} + \frac{25\pi^2}{18} +
\zeta_3 \right] \ ,
\nonumber \\
X_A & \simeq & \left[ 5 -
\frac{119\pi^2}{48} - \frac{53}{8}\zeta_3 + \frac{19}{4}\pi^2\ln2
- \frac{11\pi^4}{720} \right] + \mm \left[ -\frac{73}{8} +
\frac{41\pi^2}{8} - \frac{41\pi^4}{90} \right] \nonumber \\ \quad
& & + \mm^2 \left[ -\frac{7537}{288} + \frac{523\pi^2}{96} +
\frac{295}{32}\zeta_3 - \frac{27}{16}\pi^2\ln2 -
\frac{191\pi^4}{720} + \left( \frac{115}{48} - \frac{5\pi^2}{16}
\right) L \right] \ , \label{xa}
\nonumber \\
X_{NA} & \simeq & \left[
\frac{521}{576} + \frac{505\pi^2}{864} + \frac{9}{16}\zeta_3 -
\frac{19}{8}\pi^2\ln2 + \frac{11\pi^4}{1440} \right] + \mm \left[
\frac{91}{48} + \frac{329\pi^2}{144} - \frac{13\pi^4}{60} \right]
\nonumber \\ \quad & & + \mm^2 \left[ -\frac{12169}{576} +
\frac{2171\pi^2}{576} + \frac{377}{64}\zeta_3 +
\frac{27}{32}\pi^2\ln2 - \frac{77\pi^4}{288} + \left(
\frac{73}{16} - \frac{3\pi^2}{32} \right) L \right] \ . \label{xna}
\ea

The leading term, $\order{\omega^0}$, of these results can be
compared with the numerical estimates obtained with the zero
recoil expansions in Eq.~(14) of~\cite{Czarnecki:1998qc}; all of
our results agree within their error estimations. Our result can
also be compared with a numerical study of the top decay rate
obtained by means of Pad\'e approximations up to
$\order{\omega^2}$~\cite{Chetyrkin:1999ju}. In many cases we find
agreement. However, there are also instances where the numerical
estimates in \cite{Chetyrkin:1999ju} differ from our analytic
expressions (\ref{xna}) by a few error bar lengths, illustrating
limitations of the Pad\'e approximation in this problem. For
example, the coefficient of the $\order{\omega}$ term of the
nonabelian part $X_{NA}$ of Eq.~(\ref{xna}) is $3.3398$ whereas
the value cited in \cite{Chetyrkin:1999ju} reads $3.356(3)$,
corresponding to a $5\sigma$ discrepancy. Similarly in $X_{A}$,
the $\order{\omega}$ term is off by $3\sigma$.

\begin{figure}[htb]
\begin{tabular}{cc}
\psfig{figure=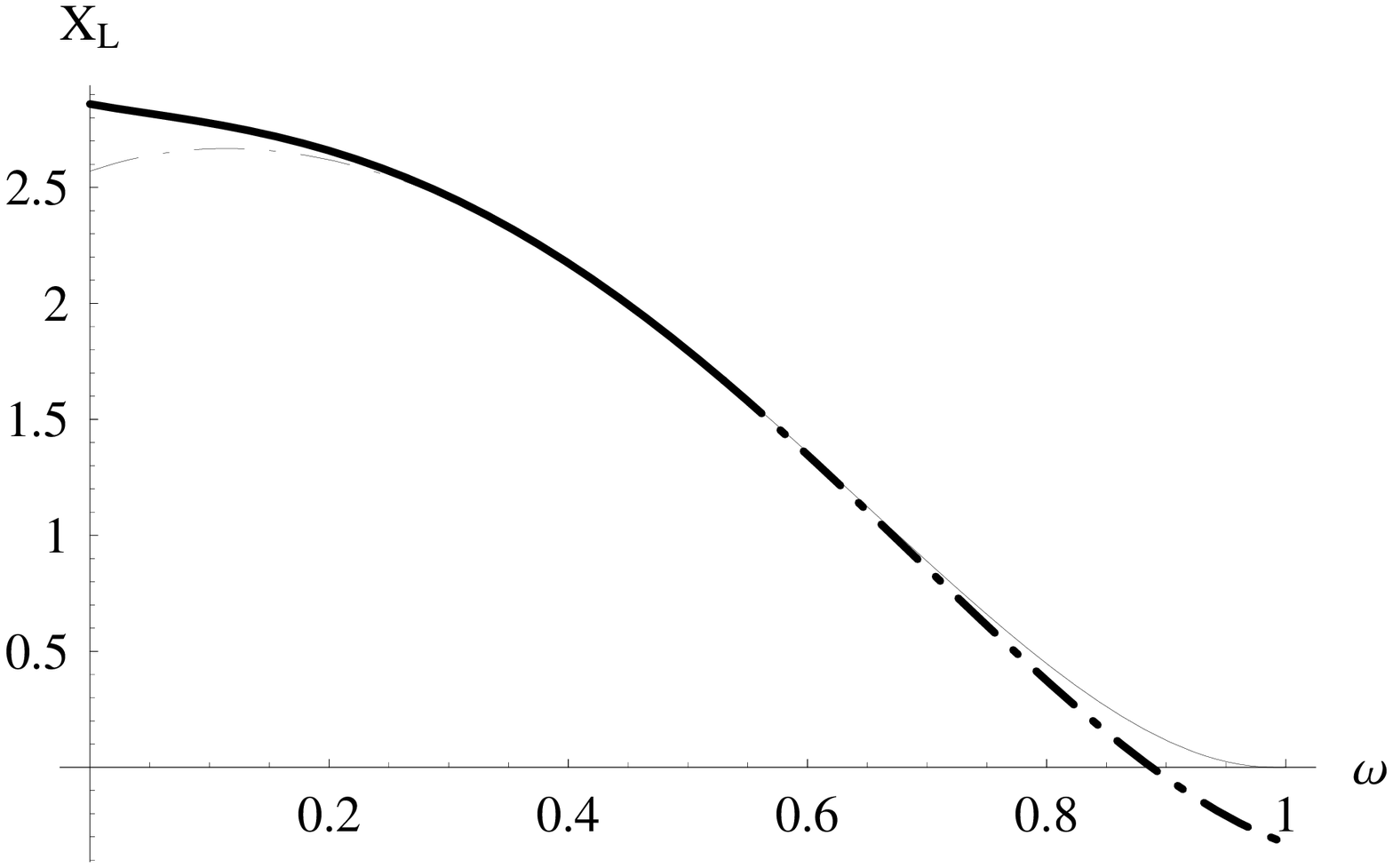,width=85mm} & \hspace{1mm}
\psfig{figure=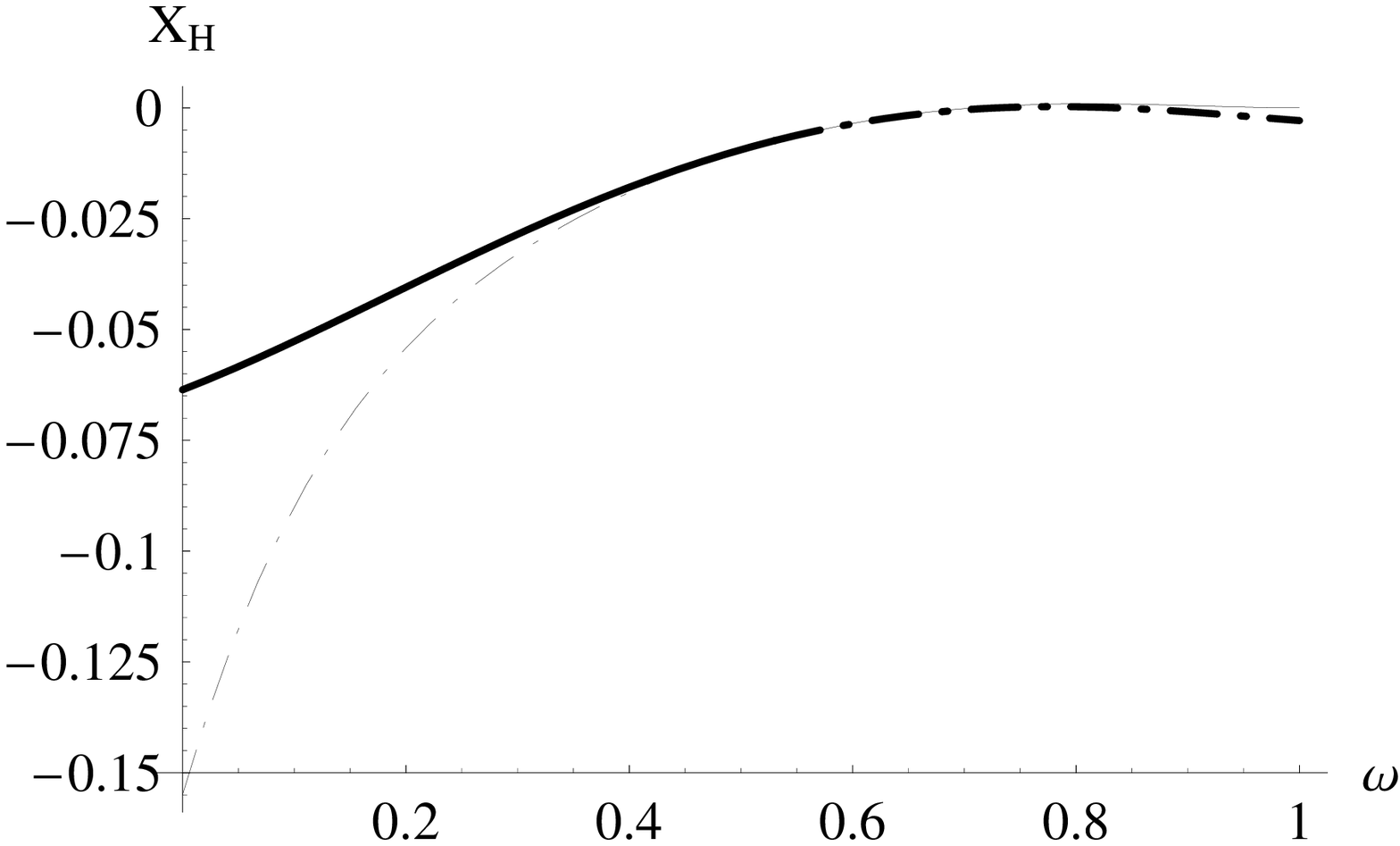,width=85mm} \hspace{5mm}
\\
\psfig{figure=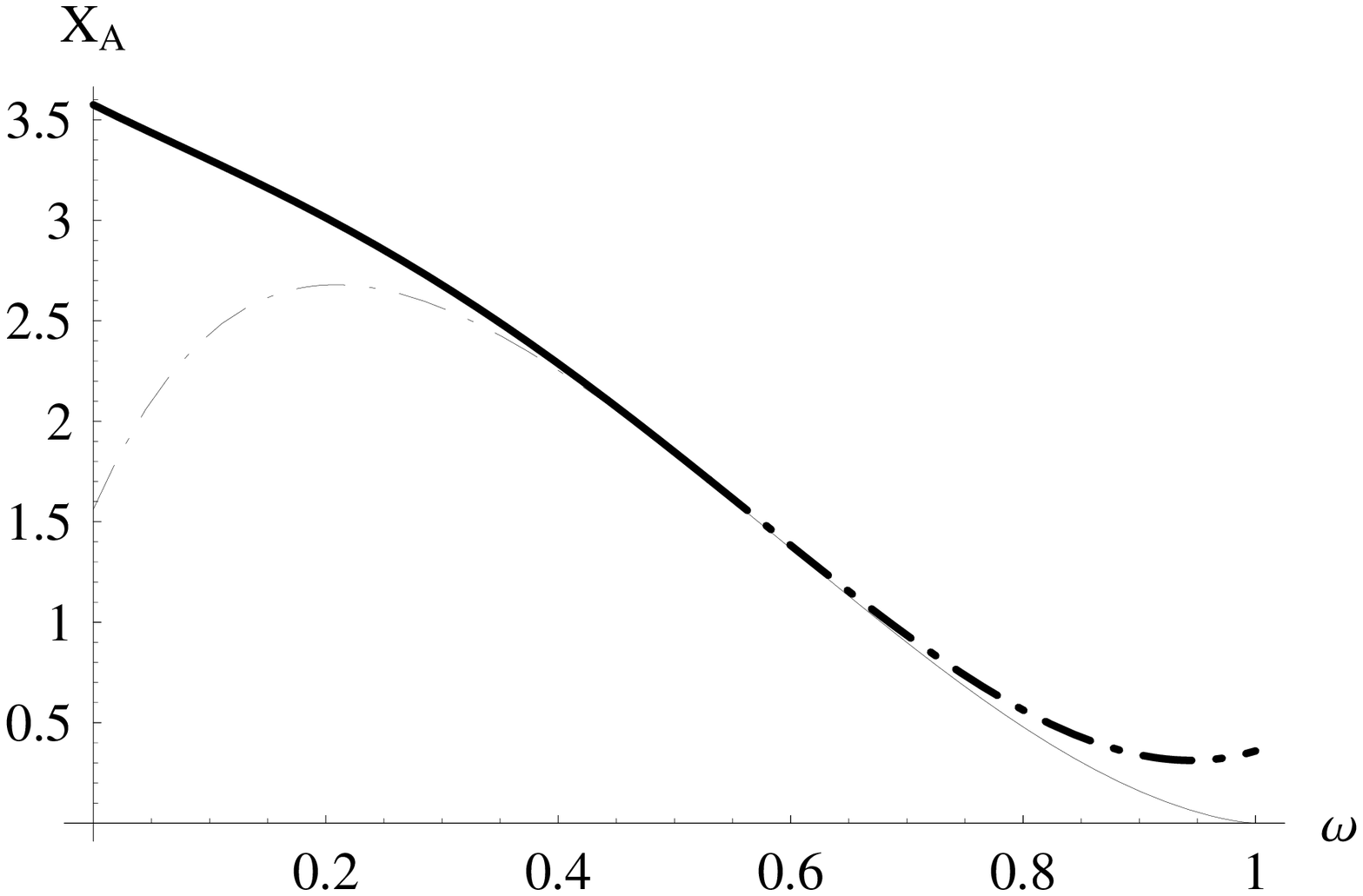,width=85mm} & \hspace{5mm}
\psfig{figure=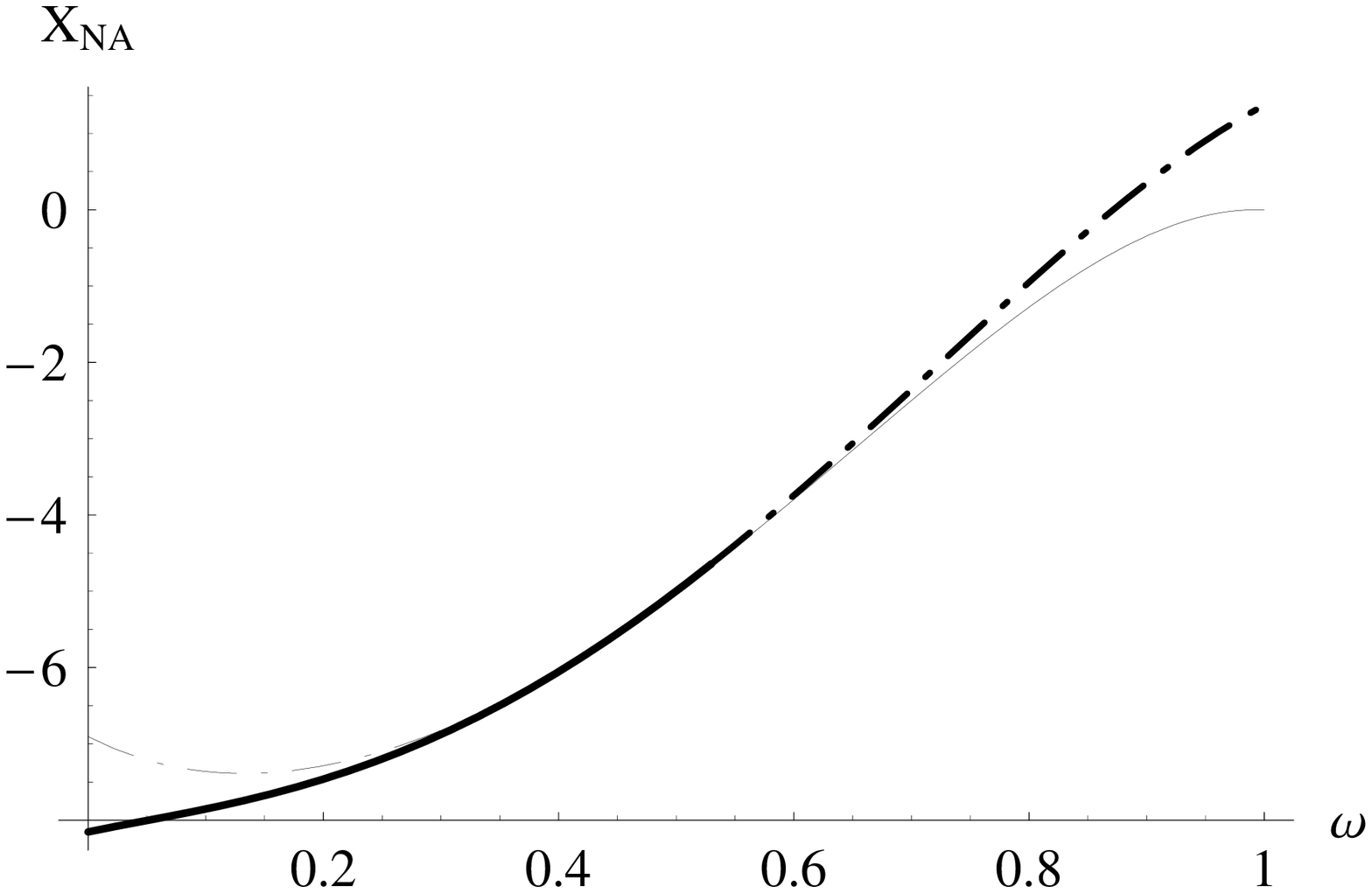,width=85mm} \hspace{5mm}

\end{tabular}
\caption{Matching of expansions around $\omega = 0$ (thick line)
and $\omega = 1$ (thin line). The solid line denotes the resulting
decay width valid in the full range of $\omega$.  Outside their
regions of validity, the expansions are shown as dash-dotted
lines.} \label{fig:matching}
\end{figure}

With a sufficient number of terms, the present expansion can be
smoothly matched with the one around the $\omega=1$ limit studied
previously~\cite{Czarnecki:2001cz} in the context of semileptonic
$b$ quark decays. The result of such a matching procedure is
depicted in the graphs in Fig.~\ref{fig:matching}. Although strict
matching of the two expansions in the entire interval $0 \leq
\omega  \leq 1$ would require a very large number of terms from
each side, a wide overlap region arises
even when only a few terms are taken into account.

The most obvious application of the above result is the precise
determination of second order QCD corrections to the top quark
decay rate. An estimation of this effect is already known, both
from numerical studies and from an extrapolation of the zero
recoil limit. However, for the measured ratio of $W$ and top
masses, $\omega  \simeq 0.213$~\cite{PDBook}, the present
expansion is the best way to calculate an accurate value of this
contribution with a reliable error estimate. Our expansion gives
$X_2 = -15.5(1)$ where the uncertainty is almost entirely due to
the experimental uncertainty of $m_t$. The theoretical error,
which originates from taking a finite number of terms in our
expansion, is $20$ times smaller and can be still easily reduced if
needed. Using $\alpha_s ( m_t) = 0.11$, we find that the two-loop
correction decreases the tree level decay rate by about $2\%$, in
agreement with earlier expectations.

Our result also provides a check of the total lifetime
calculations carried out for $\mu\to e\nu\bar\nu$ and $b\to u
l\nu$ decays. In these processes the expansion parameter $\omega$
corresponds to the invariant mass of leptons produced in the decay
and our matching procedure allows us to obtain a differential
width $d \Gamma / d \omega$ valid in the full range of $\omega$
with desired accuracy. The inclusive semileptonic decay rate $b\to
u l\nu$ can be calculated by integrating over $\omega$ within the
kinematical boundaries. Taking $N_L = 4$ and $N_H = 1$, we end up
with $\int_0^1 d \omega X_2 (\omega) = -10.644$, which almost
perfectly reproduces the $-10.648$ given
in~\cite{vanRitbergen:1999gs}. Analogously, the two-photon
correction to the muon lifetime emerges from an integration of the
abelian contribution $X_A$. We find $\int_0^1 d \omega X_A
(\omega) = 1.7797$, which is in excellent agreement with the exact
result $1.7794$ \cite{vanRitbergen:1998yd}.

To summarize, we have presented a new analytic
$\order{\alpha_s^2}$ result for the decay $t\rightarrow bW$ in
terms of a parameter $\omega=m_W^2/m_t^2$ and in the limit of
$m_b=0$, corresponding to the last remaining kinematic region in
which the $\order{\alpha_s^2}$ heavy quark decay rates were not
analytically known.  This result has enabled us to confirm or
modify slightly the corresponding results of previous numerical
calculations.  Our formulas are readily applicable to other
physical processes such as muon decay and the semileptonic
$b$ quark decay $b\rightarrow u l\nu$.

Our results depend on the imaginary parts of a novel class of
three-loop integrals, which we have obtained using two independent
paradigms for the solution of large systems of recurrence
relations.  To the best of our knowledge, this is the first time
that both approaches have been used simultaneously to obtain a new
result, and an objective analysis of the strengths and weaknesses
of each approach will increase the efficiency of other large
calculations in the future.  This augurs well for the increasingly
difficult physical problems that lie ahead.  In particular, the
top quark decay problem considered here has laid the foundation
for $\order{\alpha_s^2}$ perturbative calculations of mixing
processes such us $B_d\leftrightarrow \overline B_d$ and
$B_s\leftrightarrow \overline B_s$. Since the recently found
$\order{\alpha_s}$ effects are large and suffer from strong scale
dependence, such improvement will help use those processes as a
probe for new physics.

\emph{Acknowledgements:} We are grateful to K. Melnikov for
sharing his experience with solving large systems of recurrence
relations, and to J. Bl\"umlein and S. Moch for help with harmonic
polylogarithms, which were very helpful in determining master
integrals. This research was supported by the Science and
Engineering Research Canada, Alberta Ingenuity, and by the
Collaborative Linkage Grant PST.CLG.977761 from the NATO Science
Programme.


\begin{thebibliography}{17}
\bibitem[{csc()}]{csc}
\bibinfo{note}{Our calculations are performed with facilities of the Centre for
  Symbolic Computation at the University of Alberta}.
\end{thebibliography}
\end{document}